\documentclass[]{wlscirep}
\usepackage{physics}

\title{Optimizing a quantum reservoir computer for time series prediction}

\author[1,2,*]{Aki Kutvonen}
\author[3,+]{Keisuke Fujii}
\author[1,+]{Takahiro Sagawa}

\affil[1]{Department of Applied Physics, The University of Tokyo, 7-3-1 Hongo, Bunkyo-ku, Tokyo 113-8656, Japan}
\affil[2]{COMP Center of Excellence, Department of Applied Physics,
Aalto University School of Science, P.O. Box 11000, FI-00076 Aalto, Espoo, Finland}
\affil[3]{Graduate School of Science, Kyoto University, Sakyo-ku, Kyoto, 606-8502, Japan}
	
\affil[*]{aki.kutvonen@gmail.com}
\affil[+]{these authors contributed equally to this work}


\begin{abstract}
Quantum computing and neural networks show great promise for the future of information processing. In this paper we study a quantum reservoir computer (QRC), a framework harnessing quantum dynamics and designed for fast and efficient solving of temporal machine learning tasks such as speech recognition, time series prediction and natural language processing. Specifically, we study memory capacity and accuracy of a quantum reservoir computer based on the fully connected transverse field Ising model by investigating different forms of inter-spin interactions and computing timescales. We show that variation in inter-spin interactions leads to a better memory capacity in general, by engineering the type of interactions the capacity can be greatly enhanced and there exists an optimal timescale at which the capacity is maximized. To connect computational capabilities to physical properties of the underlaying system, we also study the out-of-time-ordered correlator and find that its faster decay implies a more accurate memory. Furthermore, as an example application on real world data, we use QRC to predict stock values.

\end{abstract}
\begin{document}

\flushbottom
\maketitle

\thispagestyle{empty}

\section*{Introduction}

Partly due to the increase in available data and computational speed, machine learning models based on neural and deep neural network architectures have recently been remarkably successful\cite{ANN1,ANN2,ANN3,ANN4}. Recurrent neural networks, which are designed for modeling problems with temporal aspects, have been successful in various end-to-end systems such as machine translation, speech interpretation/generation and time series prediction \cite{RNN1,RNN2,RNN3}. Traditionally the neural networks have been trained using back propagation through time algorithm, in which the network is trained against loss function by gradient descent in parameter space \cite{backpropagation}.

In this paper we consider an alternative way of utilizing recurrent neural networks, the framework of reservoir computing \cite{RC1,RC2,RC3,RC4}. Unlike in the common back propagation approach, in reservoir computing the recurrent part of the network (reservoir) is fixed during the training and only the readout weights are optimized. The reservoir networks can be directly implemented to physical systems allowing extremely fast state-of-the-art information processing \cite{ARC1,ARC2,ARC4,ARC5,ARC6,ARC7}. Here we consider a quantum reservoir computer, where the network dynamics are governed by rich quantum mechanical dynamics. Specifically, we consider a model based on fully connected transverse Ising model, introduced in Ref. \cite{Fujii2017}.  While quantum reservoir computing has demonstrated its computational capability in principle, it is still unknown how to enhance its computational power by engineering the underlying physical properties. Here we focus on engineering the variation of magnetic coupling, inter-spin interaction and input intervals. The results provide useful information in how future networks should be configured for use in reservoir computing applications. In this work, we engineer quantum reservoirs which allow up to 50$\%$ memory performance boost compared to the previous setups. We note that the aim is not to build an universal quantum computer and run algorithms on it, rather we are interested in a solver for recurrent neural networks which takes the advantage of rich quantum mechanical dynamics and exponentially scaling Hilbert spaces. 

One of key attributes of recurrent neural networks is the memory performance which determines the applicability of the network for various tasks \cite{STM}. We measure the memory performance with the long short term memory test and show that the memory accuracy and length of a quantum reservoir computer can be optimized by choosing an optimal timescale for the process or altering the inter-spin interactions. In addition, to better understand the underlying dynamics and its relation to the memory performance, we investigate the speed of  information encoding from the inputs to the internal state of the quantum reservoir computer by using the out-of-time-ordered correlator (OTOC)\cite{otoc}. 

\section*{Setup}

\begin{figure}{}
\centering
    \includegraphics[width=0.33\textwidth]{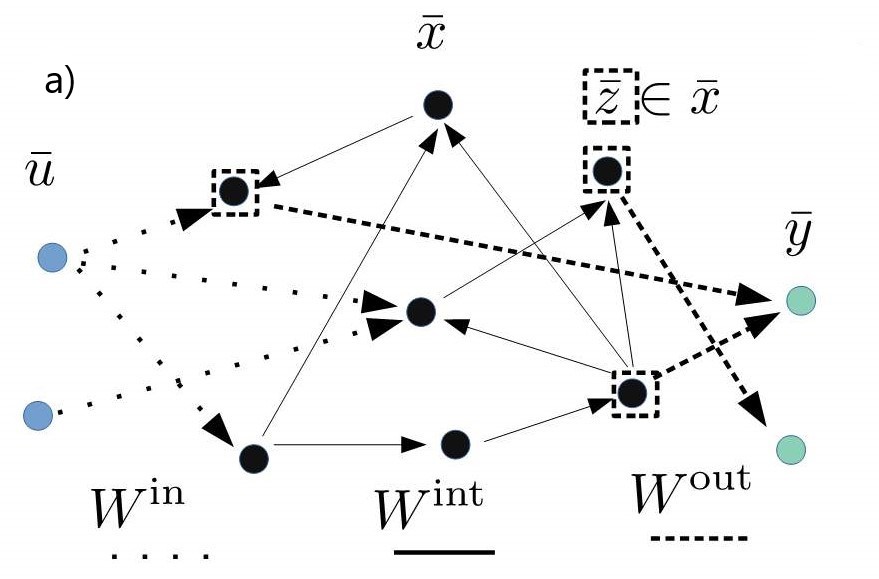}
       \includegraphics[width=0.4\textwidth]{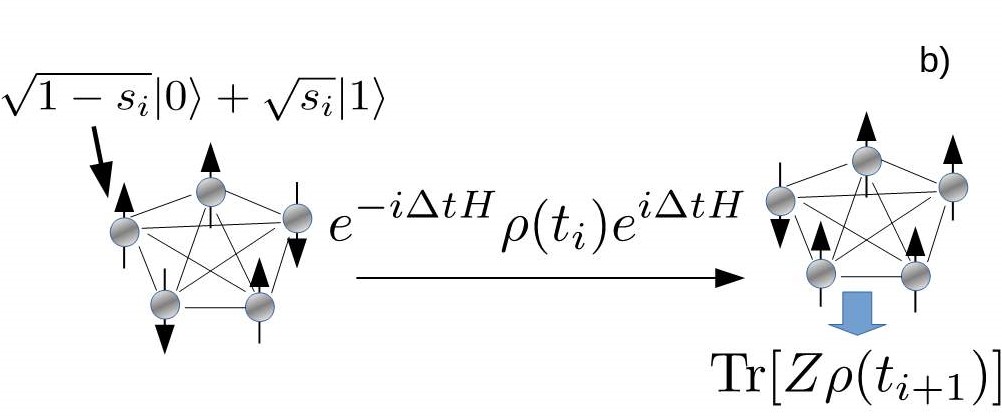}
    \caption{a) In reservoir computing the input data is fed from the input nodes $\bar u$ with weights $W^{\textbf{in}}$ to the network hidden nodes $\bar x$. The internal immutable dynamics are governed by $W^{\textbf{in}}$. Outputs are obtained by weighting the readout nodes $\bar z$, corresponding to  $\langle Z_i \rangle=\text{Tr}[Z_i \rho]$ in our setup, with the readout weights $W^{\textbf{out}}$. b) In quantum reservoir computing an input $s_i$ is fed to a spin by setting its state to $|\Psi_{s_i} \rangle = \sqrt{1-s_i}|0\rangle + \sqrt{s_i} | 1 \rangle$. After free time evolution the readout node values are obtained as the ensemble averages of the spins $\langle Z \rangle$.}
    \label{fig1}
\end{figure}

We start by describing the standard protocol of recurrent neural networks and reservoir computing. Given the discrete splitting of time $\{t_i\}$, values of the input nodes $\bar u (t_i)=\{u_j(t_i)\}_{j=0}^{n_u}$, the network nodes $\bar x (t_i)=\{x_j(t_i)\}_{j=0}^{n_x}$ and the output nodes $\bar y (_it)=\{y_j(t_i)\}_{j=0}^{n_y}$, the update equation for a simple recurrent neural network is given by

\begin{equation}
x_i(t_{i+1})=f[\sum_j^{n_u} W^{\textbf{in}}_{x_i,u_j}u_j(t_{i+1})+ \sum_j^{n_x} W^{\textbf{int}}_{x_i, x_j}x_j(t_i)], 
\label{eq:xupdate}
\end{equation}
where $f$ is some nonlinear differentiable function such as sigmoid function, $W^{\textbf{in}}$, $W^{\textbf{int}}$ are the transition weights from the input nodes and hidden nodes, and $n_u$, $n_x$ and $n_y$ are the number of input, hidden and output nodes, respectively. A simple recurrent neural network is depicted in Fig. \ref{fig1} (a). The network outputs are given by
\begin{equation}
y_i(t_{i+1})=\sum_j^{n_{out}} W^{\textbf{out}}_{y_i,z_j}z_j(t_i)  ,
\label{eq:yupdate}
\end{equation}
where we assume that the readout signals are gathered from a subset $\{z_i\} \in \{x_i\} $ of size $n_{out}< n_x$ from hidden nodes. In general the output $y_i(t_{i+1})$ could have contribution from last inputs $\bar u(t_i)$ and outputs $\bar y(t_i)$ or the nodes itself could have more complex design \cite{LSTM}. \\

Given the values of input nodes $\bar u (t_i)$ (input sequence), the training goal is to find a set of weights $W^\textbf{out}$, which minimize the error $\mathcal{E}=\sum_{i} \| \bar y^{\textbf{targ}}(t_i)-\bar y(t_i) \|$, where $y^{\textbf{targ}}(t)$ is the target sequence and $\| \dot \|$ denotes a differentiable distance function, such as the mean square distance or cross entropy between distributions. The standard method of back propagation for optimizing typically millions of parameters is to evaluate and numerically descend along the gradients $d \mathcal{E} / dW$ for all elements in $W^{\textbf{in}}$ and $W^{\textbf{int}}$. Reservoir computing seeks to optimize the task with less computational power by training the output weights $W^{\textbf{out}}$ alone. In the case of the mean square error the task reduces to the standard linear regression task, and the optimal configuration of weights, $W^\textbf{out}_{opt}=\arg \min_{W^\textbf{out}}  \sum_{i} ( \bar y^{\textbf{targ}}(t_i)-\bar y(t_i))^2$, is given by
\begin{equation}
W^\textbf{out}_{opt}=\mathcal{Z}^{-1}\mathcal{Y} ,
\label{eq:Wout}
\end{equation}
where $\mathcal{Y}_{i,j}=y^\textbf{targ}_j(t_i)$ and $\mathcal{Z}^{-1}$ denotes the Moore–Penrose pseudo inverse of matrix $\mathcal{Z}_{i,j}=z_j(t_i)$. Since in reservoir computing the inter network optimization is not necessary during the training, a physical system can drive the network dynamics directly.

The memory of the network is measured by the short term memory (STM) task defined as a task to remember the input $d$ steps ago:
\begin{equation}
y^{\textbf{targ}}_{d}(t_i)=s_{i-d}.
\label{eq:ytarg}
\end{equation}
The memory accuracy is defined as the correlation between the prediction and the target function:
\begin{equation}
\mathcal R (d)=\text{cov}^2(y,y^{\textbf{targ}}_d) / \sigma^2(y)\sigma^2(y^{\textbf{targ}}_d),
\label{eq:R}
\end{equation} 
where $\text{cov}$ denotes covariance and $\sigma^2$ variance. The memory capacity $\mathcal C$ is defined as $\mathcal C=\sum_{d=1}^{d_c} \mathcal R(d)$, where $d_c$ is a cut off distance, which we set to 100.

For efficient usage, reservoir computing requires complex dynamics, large network sizes and bounded dynamics. These requirements are naturally met by interacting quantum mechanical spin systems, whose state space scales exponentially with respect to the number of spins and the dynamics are governed by complex unitary dynamics. The notation here follows the standard quantum mechanics notation. The state of a spin is a two-dimensional complex vector spanned by the eigenstates of the Pauli $Z$ operator, $\{ | 0 \rangle , | 1 \rangle \}$. The total state space of $\it N$ qubits is a tensor product space of the two-dimensional individual spin spaces. The total (pure) state of the system is represented by a $2^N$-dimensional vector $| \Psi \rangle$. A statistical mixture of pure states can be described by $2^N \times 2^N$-dimensional density matrix $\rho$. For a closed quantum system, the time evolution is governed by the time independent Schr\"odinger equation, which for the density matrix can be written as

\begin{equation}
\rho(t+\Delta t)=e^{-i\Delta tH}\rho(t)e^{i \Delta tH} ,
\label{eq:rhot}
\end{equation}
 where $H$ is a $2^N \times 2^N$-dimensional hermitian matrix, the Hamiltonian of the system, which defines the system dynamics. Here we consider an extensively studied model, the fully connected transverse field Ising model, whose Hamiltonian is given by 
\begin{equation}
H=\sum_{i,j} J_{i,j} X_i X_j + h_i Z_i,
\label{eq:H}
\end{equation}
, where $X_i$ and $Z_i$ are the Pauli $X$ and $Z$ operators at site $i$. In this model all the spins interact with each other in $x$-direction and are coupled to an external magnetic field in z-direction.

We consider one-dimensional inputs $\bar s(t_i)=s_i$ and outputs $\bar y(t_i)=y_i$. At time $t_i$ the input $s_i$ is fed to the system by setting the state of the first spin to $\ket{ \Psi_{s_i} } =\sqrt{1-s_i}\ket 0+\sqrt{s_i}\ket 1$. The density matrix of the system then becomes $\rho \mapsto \ket{ \Psi_{s_i}} \bra{ \Psi_{s_i} }  \bigotimes \text{Tr} _1 [\rho] $, where $\text{Tr}_1$ denotes the trace over the first spin degree of freedom. 

After the input is set, the system continues evolving itself for time $\Delta t$. During this time the dynamics are governed by Eq. \eqref{eq:rhot} and the information encoded in the first spin will spread through the system. During $[t_i,t_{i+1}]$ we measure the average spin values in the $z$-direction, $\langle Z_i \rangle=\text{Tr}[Z_i \rho]$, which corresponds to the readout nodes in our setup. After gathering a sufficient amount of input-output signal pairs, the readout weights $W^\textbf{out}$ are trained according to Eq. \eqref{eq:Wout} and the corresponding outputs and the future predicted outputs are obtained from Eq. \eqref{eq:yupdate}. The input-readout loop is illustrated in Fig. \ref{fig1} (b).

Since the number of spins in our setup, $N_{s}$, is fairly small, we also gather the readout signals from intermediate times between $[t,t+\Delta t]$. The input timescale $\Delta t$ is divided into $N_v$ time steps and thus the intermediate times are given by $t_i^k=t_i+k \Delta t /N_v$, where $k=1,..,N_v$. The readout signals are gathered as $\text{Tr}[Z_j \rho(t_i^k)]$ for all the spins $j=1,..,N_{s}$ and the intermediate times $t_i^k$. Thus the total number of the output signals is $N_{spins}N_v$.

We note that an experimental realization of the setup requires multiple copies of the same system in order to make the effect of measurement back-action irrelevant. This is feasible by using nuclear magnetic resonance systems, where a huge ensemble of identical molecules can be utilized. Further details on the input-readout loop with practical implementation considerations are available in Ref. \cite{Fujii2017}.

\section*{Results}

We set the number of spins and intermediate times to $N_{s}=6$  and $N_v=10$, respectively, which gives a relatively good performance at reasonable computational cost \cite{Fujii2017}. The input sequences were random binary inputs, $s_i=\{0,1\}$, sampled for each simulation separately. Initial state was set to the maximally entangled state, where all the possible states $|\Psi\rangle$ appear with the same probability $1/2^{N_{s}}$. All of the simulations consisted of an initial equilibration phase of 2000 time steps during which the inputs were injected but no training was done. During following 3000 time steps the input and readout signals were gathered and the output weights were trained according to Eq. \eqref{eq:Wout}. The testing was done by generating a new set of 1000 data points. The network output prediction of Eq. \eqref{eq:yupdate} was then compared to the target output of Eq. \eqref{eq:ytarg}.

We considered various values of magnetic coupling $h$, inter-spin interactions $J$ (Eq. \eqref{eq:H}) and input intervals $\Delta t$. Values for both $h$ and $J$ were sampled from 0.5 centered uniform distribution $\mathcal{U}[0.5-h_s/J_s,0.5+h_s/J_s]$. The scale parameters $h_s$ and $J_s$ simulated were $h_s, J_s \in \{0.01,0.1,0.25,0.50,2,3\}$ and the number of simulations for each setup was 25. We also sampled $J$ from Beta-distribution with values $(\alpha, \beta) = \{(0.1,0.1),(0.7,0.7),(0.9,0.9),(1.1),(2,2),(9,9),(100,100)\}$ while keeping the value of $h=0.5$ fixed (see Supplementary Fig. S1 online). Figure \ref{fig2} shows the memory capacity as a function of the input timescale $\Delta t$ for selected scale parameters $h_s$ and $J_s$. For all the parameters tested, larger deviation in $J$, corresponding to larger scale values of $J_s$ and smaller values of $(\alpha, \beta)$, resulted in larger maximum capacity. However, the given boost in performance saturated around $J_s=0.5$ and $(\alpha, \beta)=(1,1)$, corresponding to uniform $[0,1]$ sampling. Within the parameters sampled, the values of $h$ did not have considerable effect on the memory capacity.

In all setups there existed an optimal input timescale $\Delta t_{\text{opt}}$ at which the memory capacity is maximized as shown in Fig. \ref{fig2}. The STM task measures how well the past input signals can be decoded from readout signals $\{\langle Z_i\rangle\}_{i=1}^6$ during time window $[t_i,t_{i+1}]=[t_i,t_i+\Delta t]$, using the simple linear map of Eq. \eqref{eq:yupdate}. On too short timescales the dynamics are almost linear at $t \in [t_i,t_{i+1}]$ and thus there is no prediction power outside the linear signal regime. On too long timescales the signals are too chaotic during $t\in [t_i,t_{i+1}]$ giving rise to sub-optimal performance. Thus at optimal timescales the dynamics are not completely linear but not yet too chaotic either. Signals at different timescales are illustrated in the lower panel of Fig. \ref{fig2}.

\begin{figure}{}
\centering
    \includegraphics[width=0.55\textwidth]{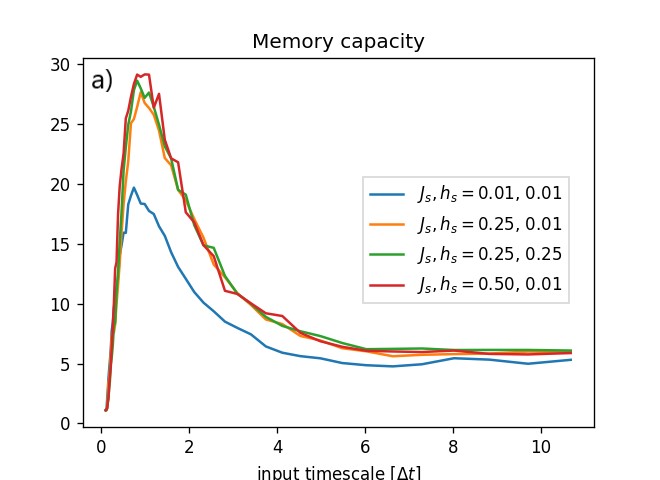}
       \includegraphics[width=0.55\textwidth]{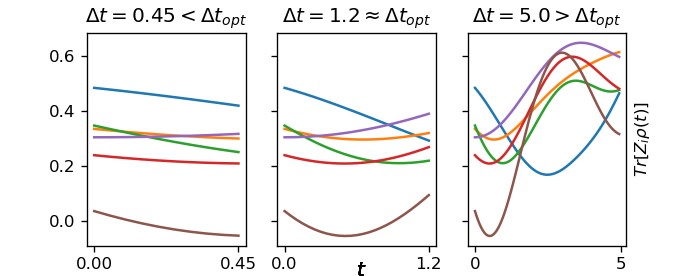}
    \caption{a) Memory capacity shown as a function of the input timescale $\Delta t$. The values for $h$ and $J$ are sampled from $\mathcal{U}[0.5-h_s,0.5+h_s]$ and $\mathcal{U}[0.5-J_s,0.5+J_s]$, respectively. Deviation in $J$ leads to better performance but the deviation in $h$ is less important. Furthermore, for all parameters simulated, there exists an optimal timescale for $\Delta t$, which maximizes the capacity. b) Readout signals Tr$[Z_i \rho(t)]$ for three different time windows $[t_i,t_i+\Delta t]$ for parameters $h_s=0.01$, $J_s=0.5$. The leftmost, center and the rightmost panels corresponds to timescales shorter, equal and longer than the optimal timescale $\Delta t_{\text{opt}}$, respectively.}
    \label{fig2}
\end{figure}

\begin{figure}{}
\centering
    \includegraphics[width=0.44\textwidth]{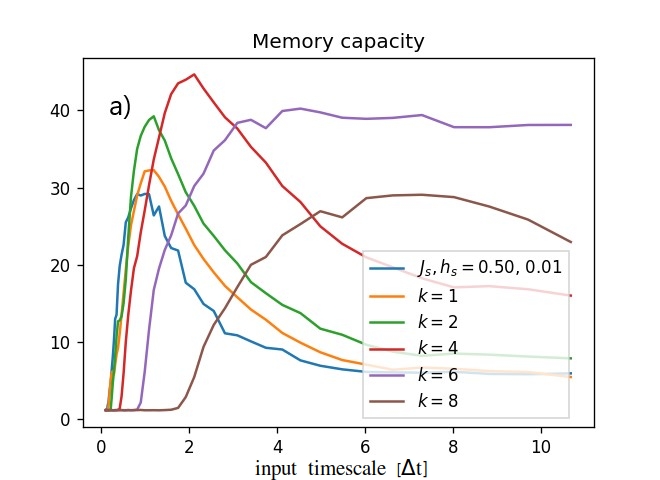}
         \includegraphics[width=0.24\textwidth]{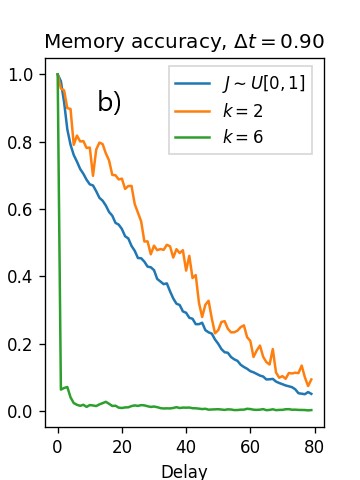}
       \includegraphics[width=0.24\textwidth]{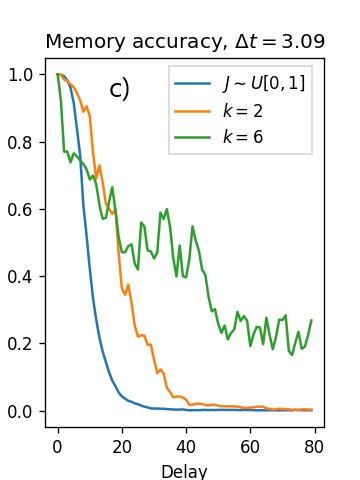}
    \caption{a) Memory capacity for different inter-spin couplings $J^k_{i,j}=(i+j)^k/c_k$. The maximum capacity has a non-monotonic behaviour in $k$, while the location of the maximum shifts towards larger $\Delta t$ in increasing $k$. b) Memory accuracy at the short time interval. The interval $\Delta$ is too short for $k=6$ while the other setups exhibit the long but inaccurate memory. c) At long time intervals the setups exhibit the short but accurate memory. Higher $k$ values result into a longer but inaccurate of memory.
 }
    \label{fig3}
\end{figure}

Instead of sampling $J$ and $h$ from a distribution, we considered engineered forms of $J$ while keeping the value of $h$ fixed to 0.5. The coupling between the spins was set to
\begin{equation}
J^k_{i,j}=(i+j)^k/c_k,
\label{eq:J}
\end{equation}
where $i$ and $j$ are the spin indexes, $k$ is a scaling parameter and $c_k=\sum_{i,j}2N_s^2(i+j)^k$ is a $k$-dependent constant which ensures comparable energy scales between setups with the mean interaction $\textbf E[J^k]=0.5$. In addition to the fixed mean, the form of $J^k$ above has the benefit of inducing deviation in $J$ with simple parametrization $k$ without additional noise induced by random sampling. Furthermore, all the spins have different average couplings to the other spins, $\sum_j J^k_{i,j} < \sum_j J^k_{i+1,j}$,  $ i \in {1,2,..,5}$. We expect that this in general results into more rich and divergent signals increasing the prediction power of Eq. \eqref{eq:yupdate}. In fact, compared to the random sampled $J$, the memory capacity can be increased up to 50\% by using $J^k$, as shown in Fig. \ref{fig3}. Maximum memory capacity has a non-monotonic dependence on $k$ and the best performance is obtained with $k=4$. Furthermore, the larger the $k$, the larger the optimal input interval $\Delta t_{\text{opt}}$. The number of simulations was 10 for each of the value $k \in \{1,2,4,6,8\}$.

For practical purposes, the memory accuracy $\mathcal R(d)$ (Eq. \eqref{eq:R}) may be more important than the overall capacity.  Figure \ref{fig3} shows the memory accuracy as a function of the delay $d$ for selected values of $k$ and input intervals $\Delta t$. On short $\Delta t$ the memory extends up to delays 100 but is inaccurate throughout the delay range. On the contrary at long intervals $\Delta t$ the memory is shorter but more accurate. As the capacity $\mathcal C=\sum_d \mathcal R(d)$ measures the sum of accuracies over memory length, the largest $\mathcal C$ is obtained somewhere in the intermediate timescales where the correlation is relatively good over all the delays up to the cutoff delay.

It is also interesting to compare the memory type to the signals given in the lower panel of Fig \ref{fig2}. On longer intervals $\Delta t$ the time window seems to give a good view of the last couple of time windows but the prediction power cannot extend far due to the chaotic nature of the signals, corresponding to the short but accurate memory in the STM test. Vice versa, at shorter intervals $\Delta t$ the linear prediction extends longer in the past but it is never quite accurate. 

When comparing the different $J^k$ at the same input intervals, the larger $k$ exhibits the longer but less accurate memory and smaller $k$ values are associated with the shorter but more accurate memory. Thus, like in the case of the input interval $\Delta t$, also the intermediate $k$ value results in the highest memory capacity (Fig. \ref{fig3}). We note that the short but accurate type of memory associated with timescales $\Delta t > \Delta t_\text{opt}$ and and lower values of $k$ might be preferred for practical applications where  a high enough accuracy is needed.

\begin{figure}{}
\centering
    \includegraphics[width=0.65\textwidth]{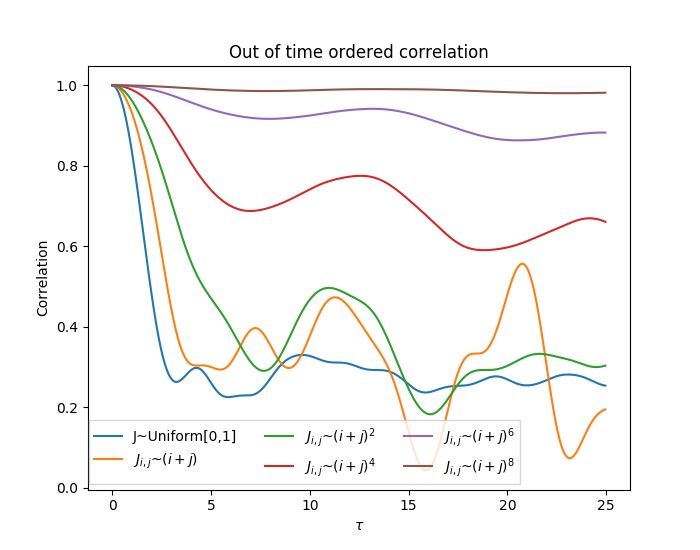}
    \caption{The OTOC between the input spin (site $i=1$) and the other spins for different couplings  $J^k_{i,j}=(i+j)^k/c_k$. The larger the $k$, the slower the information decays from the input spin. This is in agreement with the notion that the input spin is more weakly coupled
 to the other spins with higher values of $k$.}
    \label{fig4}
\end{figure}

In order to investigate the information spreading from the input spin to the other spins, we studied the out-of-time-ordered correlator (OTOC) defined as:
\begin{equation}
O(t)=\frac{1}{N_{s}} \sum_{j=2}^{N_{s}} \text{Tr}[Z_1(t_0)Z_j(t_0+t)Z_1(t_0)Z_j(t_0+t)].
\end{equation}
The OTOC measures delocalization of information and is previously used in the context of fast scrambling of black hole and thermalization of closed quantum systems \cite{otoc,Maldacena2016}. We ran 60 simulations, each consisting of an initial washout phase of 20000 time steps with time step $\Delta \tau =0.05$. Figure \ref{fig4} shows the OTOC for different $J^k$ couplings and the case that $J$ is sampled from $\mathcal U [0,1]$. The larger the value of $k$ is, the slower the OTOC decays, signaling slower decay of information from the first spin to the other spins. This is in agreement with the notion that the average interaction of spin 1 to the others, $\sum_j J_{1,j}/N_s=\sum (1+j)^k/ (c_k N_s)$, is a decreasing function in $k$. Larger values of $k$ means that the first spin is more weakly coupled to the other spins resulting into a slower decay of information.

Comparison of the memory capacity and the OTOC, Figs. \ref{fig3} and \ref{fig4}, shows that the slower decay of OTOC implies a larger optimal timescale for the STM task. Thus the system needs more time for effective distribution of information (encoding) to the other spins. These systems exhibit the long but inaccurate type of memory. The encoding takes time and the encoded information can be decoded back after a long time, but the decoding is never particularly efficient. In contrary, the random sampled and low $k$ setups are able to encode the information to the other spins fast and the decoding is efficient, but the memory does not extend as far. We can further synthesize desired memory properties by spatially multiplexing distinct quantum reservoirs \cite{multiplex}.

\begin{figure}{}
\centering
    \includegraphics[width=0.38\textwidth]{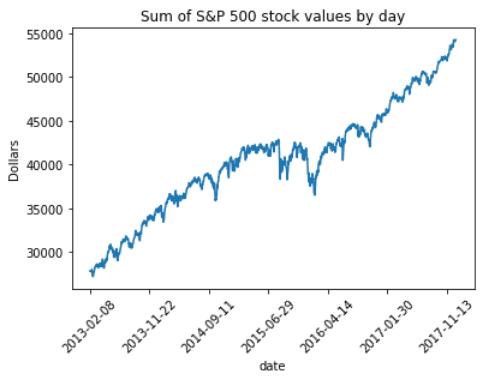}
         \includegraphics[width=0.47\textwidth]{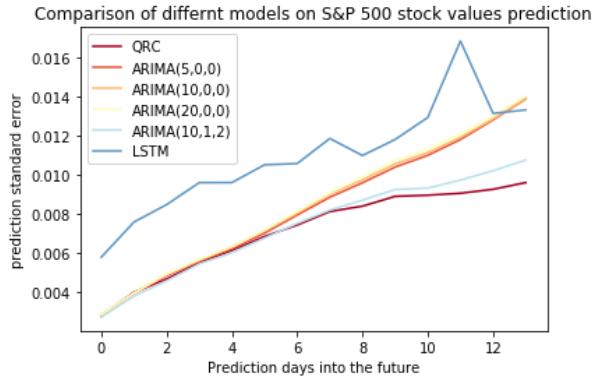}
    \caption{a) The time series data of daily closing values of S\&P 500 stocks by date. b) Mean absolute error for different prediction models as a function of how many days ahead the prediction date is. The results are averages over 100 different starting dates.}
    \label{fig5}
\end{figure}

In order to demonstrate the real world applicability of the quantum reservoir computer, we used future stock value prediction as an example task. As an example dataset we used the Standard \& Poor's (S\&P) 500 companies stock values from February 8th 2013 to December 29th 2017 \cite{stocksdata}. We took the the sum of daily closing values of the 500 companies as our daily time series data of 1232 data points shown in Fig. \ref{fig5}.

The task was to predict the time series 14 days ahead using the past data, i.e. to predict $[x_i,..,x_{i+13}]$ given $[x_1,..,x_{i-1}]$. We measured the accuracy for each future day ahead using mean squared error. Since the data is noisy and the accuracy highly depends on the starting date (time index $i$), we repeated the task for 100 different starting days $i \in [1100,..,1199]$. We compared the QRC's performance with other widely used time series forecasting methods, namely an auto regressive integrated moving average (ARIMA) model and a recurrent neural network based long short term memory (LSTM) model \cite{arima,LSTM}. The results shown in Fig. \ref{fig5} indicate that the QRC performs well compared to the other models within the selected parameters.

We did not perform an exhaustive parameter search, rather the focus was on demonstrating that the QRC can be applied to real world data. In the initial washout phase we fed the training data through the QRC once without collecting any signals. At the training phase we fed the data, collected the readout node values and trained the output matrix $W^\textbf{out}$ against the target output according to the Eq. \eqref{eq:ytarg}. We set the number of qubits to 6, inter spin interactions to Eq. \eqref{eq:J} with $k=2$ and the magnetic field was sampled from random uniform (0,1) distribution. The number of virtual nodes $N_v$ was set to 2 and we also accumulated the node values from the last 4 data injection points to our readout node values. That is to say, prediction for data points $[x_i,..,x_{i+13}]$ was done using readout node values $\text{Tr}[Z_j \rho(t_{i-k}-l \Delta t /2)]$, where $j \in [1,..,6]$, $k \in [0,..,4]$ and $l \in [0,1]$. The data input timescale was set to $\Delta t=0.6$, corresponding to the time scale 0.3 between the virtual nodes. This timescale corresponds to $\Delta t=3$ with 10 virtual nodes, a time scale which resulted into an accurate memory on shorter delays as shown in the right panel of Fig. \ref{fig3}. 

The parameters $p$, $q$ and $d$ in ARIMA$(p,d,q)$ correspond to the number of lagged time series points used for predictions, the number of differencing operations on the data and the number of lagged forecast errors used for the predictions, respectively. The model was implemented using the Python statsmodels package \cite{statsmodels}. The LSTM model consisted of 64 LSTM units, which were trained for 30 epochs using "Adam" optimizer with learning rate of 0.001, and implemented using Keras framework \cite{keras}. Furthermore, due to computational limitations, the LSTM model was not trained for each prediction starting point $x_i$ using data up to $x_{i-1}$. Instead the LSTM model was only trained once on data $[x_1,...,x_{1099}]$ using delay of 30 timesteps, which partly explains the lower accuracy.

\section*{Summary}

We studied the memory capacity and accuracy of a quantum reservoir computer based on the fully connected transverse field Ising model. We observed that the input data interval $\Delta t$ has a system dependent optimal value $\Delta t_\text{opt}$ which leads to the maximum capacity. Furthermore, our results suggest that the deviation in inter-spin interaction strengths leads to improved memory capacity up to some limit. While the maximum capacity is a sign of performance, in practical applications we may be interested in the memory accuracy, which is in general improved by using a slightly longer input time interval than $\Delta t_{\text{opt}}$. In addition, our results suggest that the fast decay of the OTOC is a sign of fast encoding of information from the input spins to the others. These setups exhibit the accurate memory at the cost of memory length. Furthermore, as a real world example application, we predicted stock prices using the quantum reservoir computer. Possible future research directions include more advanced data input-output strategies, which could implement e.g., multi-dimensional inputs and outputs, mixture of multiple readout timescales and efficient strategies for determining optimal timescales for given tasks.

\bibliography{QRC}

\section*{Acknowledgements}

A.K. is supported by Jenny and Antti Wihuri Foundation through Council of Finnish Foundations' Post Doc Pool, T.S. by JSPS KAKENHI Grant Number JP16H02211 and K.F. by JST PRESTO Grant Number JPMJPR1668, JST ERATO Grant Number JPMJER1601, and JST CREST Grant Number JPMJCR1673. We thank Eiki Iyoda for useful discussions.

\section*{Author contributions statement}

A.K. conducted the simulations and wrote the first draft of the manuscript. A.K., T.S. and K.F. analyzed the results and reviewed the manuscript.

\section*{Additional information}

The authors declare no competing interests.

\end{document}